\documentclass{PoS}

\usepackage{amsmath}

\newcommand{\be}{\begin{equation}}
\newcommand{\ee}{\end{equation}}
\newcommand{\bea}{\begin{eqnarray}}
\newcommand{\eea}{\end{eqnarray}}
\newcommand{\bt}{\begin{tabbing}}
\newcommand{\et}{\end{tabbing}}
\newcommand{\bi}{\begin{itemize}}
\newcommand{\ei}{\end{itemize}}
\newcommand{\ben}{\begin{enumerate}}
\newcommand{\een}{\end{enumerate}}
\newcommand{\nn}{\nonumber}

\newcommand{\calO}{{\mathcal O}}
\newcommand{\bfp}{{\bf p}}
\newcommand{\bfx}{{\bf x}}
\newcommand{\bfr}{{\bf r}}
\newcommand{\crad}{\langle r^2 \rangle}

\title{
   \begin{picture}(0,0)(0,0)%
   \put(355,75){\makebox(0,0)[l]{\textnormal{\normalsize KEK-CP-216}}}%
   \end{picture}%
   Pion vector and scalar form factors with dynamical overlap quarks
}

\ShortTitle{
   Pion vector and scalar form factors with dynamical overlap quarks
}

\author{ 
   JLQCD and TWQCD collaborations: 
   \speaker{T.~Kaneko}$^{a,b}$\thanks{E-mail: takashi.kaneko@kek.jp}, 
   S.~Aoki$^{c,d}$, 
   T.~W.~Chiu$^{e}$, 
   H.~Fukaya$^{a,f}$, 
   S.~Hashimoto$^{a,b}$, 
   T.~H.~Hsieh$^g$,
   H.~Matsufuru$^{a}$, 
   J.~Noaki$^{a}$, 
   T.~Onogi$^{h}$,  
   E.~Shintani$^{a}$
   and 
   N.~Yamada$^{a,b}$
   \\
   \\
   \\
   \llap{$^a$}
   High Energy Accelerator Research Organization (KEK),
   Ibaraki 305-0801, Japan 
   \\
   \llap{$^b$}
   School of High Energy Accelerator Science,
   The Graduate University for Advanced Studies (Sokendai),
   Ibaraki 305-0801, Japan
   \\ 
   \llap{$^c$}
   Graduate School of Pure and Applied Sciences, 
   University of Tsukuba, Ibaraki 305-8571, Japan
   \\
   \llap{$^d$}
   Riken BNL Research Center, 
   Brookhaven National Laboratory, Upton, New York 11973, USA
   \\
   \llap{$^e$}
   Physics Department, Center for Theoretical Sciences,
   and Center for Quantum Science and Engineering,
   National Taiwan University, Taipei, 10617, Taiwan
   \\
   \llap{$^f$}
   The Niels Bohr Institute,
   The Niels Bohr International Academy,
   Blegdamsvej 17 DK-2100 Copenhagen {\O}, Denmark
   \\
   \llap{$^g$}
   Research Center for Applied Sciences,
   Academia Sinica, Taipei 115, Taiwan
   \\
   \llap{$^h$}
   Yukawa Institute for Theoretical Physics, Kyoto University,
   Kyoto 606-8502, Japan
}

\abstract{
We calculate the pion vector and scalar form factors in two-flavor QCD.
Gauge configurations are generated with dynamical overlap quarks 
on a $16^3 \times 32$ lattice at a lattice spacing of 0.12 fm 
with sea quark masses down to a sixth of the physical strange quark mass. 
Contributions of disconnected diagrams to the scalar form factor
is calculated employing the all-to-all quark propagators.
We present a detailed comparison of the vector and scalar radii
with chiral perturbation theory to two loops.
}

\FullConference{
   The XXVI International Symposium on Lattice Field Theory\\
   July 14-19 2008\\
   Williamsburg, Virginia, USA
}

\begin{document}


\section{Introduction}
\vspace{-3mm}


Pion electromagnetic form factor $F_V(q^2)$ is one of the fundamental 
observables in hadron physics.
An analysis of experimental data based on chiral perturbation theory (ChPT) 
at two loops leads to a precise estimate of the charge radius $\crad_V$
\cite{PFF_V:ChPT:NNLO}.
A detailed comparison of $\crad_V$ between ChPT and non-perturbative 
calculations on the lattice may provide a good testing ground for 
recent lattice simulations in the chiral regime 
as well as a better understanding of the chiral behavior of $F_V(q^2)$.
 
While there is no experimental processes directly related to 
the scalar form factor $F_S(q^2)$,
the chiral behavior of the scalar radius $\crad_S$ is interesting,
as it provides a determination of the LEC $l_4$ 
and has a 6~times enhanced chiral logarithm compared to $\crad_V$.
A non-perturbative determination on the lattice is challenging,
because we need to evaluate disconnected three-point functions.

In this article, 
we update our analysis of $F_V(q^2)$ reported at the last conference
\cite{PFF:Nf2:RG+Ovr:JLQCD} with doubled statistics,
and present newly obtained results for $F_S(q^2)$.
These quantities are measured on gauge configurations 
of two-flavor QCD on a $16^3 \! \times \! 32$ lattice
generated with the overlap quark action along the fixed topology strategy
\cite{fixedQ}.
The lattice spacing determined from the Sommer scale $r_0\!=\!0.49$~fm is 
$a\!=\!0.1184(21)$~fm.
We refer the reader to Refs.\cite{Prod_Run:JLQCD:Nf2:RG+Ovr,Lat07:JLQCD:Matsufuru+Lat08:JLQCD:Hashimoto} for detailed setup and overviews of our production simulations.


\vspace{-3mm}
\section{Measurement of pion correlation functions} 
\vspace{-3mm}


We measure pion correlators through all-to-all quark propagators \cite{A2A}.
Contributions of 100 low-lying modes $(\lambda^{(k)},\,u^{(k)})$ 
$(k\!=\!1,...,N_{\rm ep};N_{\rm ep}\!=\!100)$ of the overlap operator $D$ are evaluated exactly, 
whereas the remaining high modes are taken into account stochastically
by the $Z_2$ noise method.
We prepare a single noise vector for each configuration, and 
dilute~\cite{A2A} it into $N_d = 3 \times 4 \times N_t/2$ vectors 
$\eta^{(k)}$ $(k\!=\!1,...,N_d)$ 
with support on a single value for color and spinor 
indices and at two time-slices.
The all-to-all propagator can then be expressed in a simple form
$D^{-1} = \sum_{k=1}^{N_{\rm vec}} v^{(k)}\,w^{(k)\dagger}$
$(N_{\rm vec}=N_{\rm ep}+N_d)$
with two set of vectors 
\bea
   v^{(k)} 
   & = & 
   \left\{
      \frac{u^{(1)}}{\lambda^{(1)}},
      \ldots,
      \frac{u^{(N_{\rm ep})}}{\lambda^{(N_{\rm ep})}},
      x^{(1)}, \dots, x^{(N_d)}
   \right\},
   \hspace{3mm}
   w^{(k)} 
   = 
   \left\{
      u^{(1)}, \ldots, u^{(N_{\rm ep})},
      \eta^{(1)}, \dots, \eta^{(N_d)}
   \right\},
   \label{eqn:meas:vw_vectors}
\eea
where $x^{(d)}\!=\!D^{-1}(1-\sum_k u^{(k)}\,u^{(k)\dagger})\,\eta^{(d)}$.

\begin{figure}[t]
\begin{center}
\includegraphics[angle=0,width=0.3\linewidth,clip]%
                {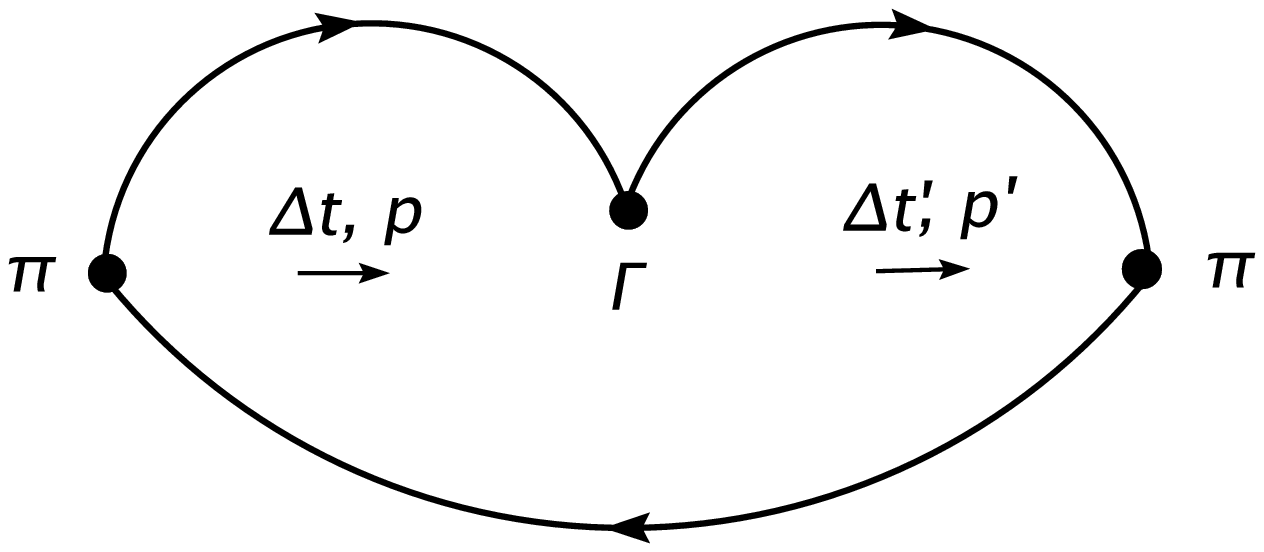}
\hspace{5mm}
\includegraphics[angle=0,width=0.3\linewidth,clip]%
                {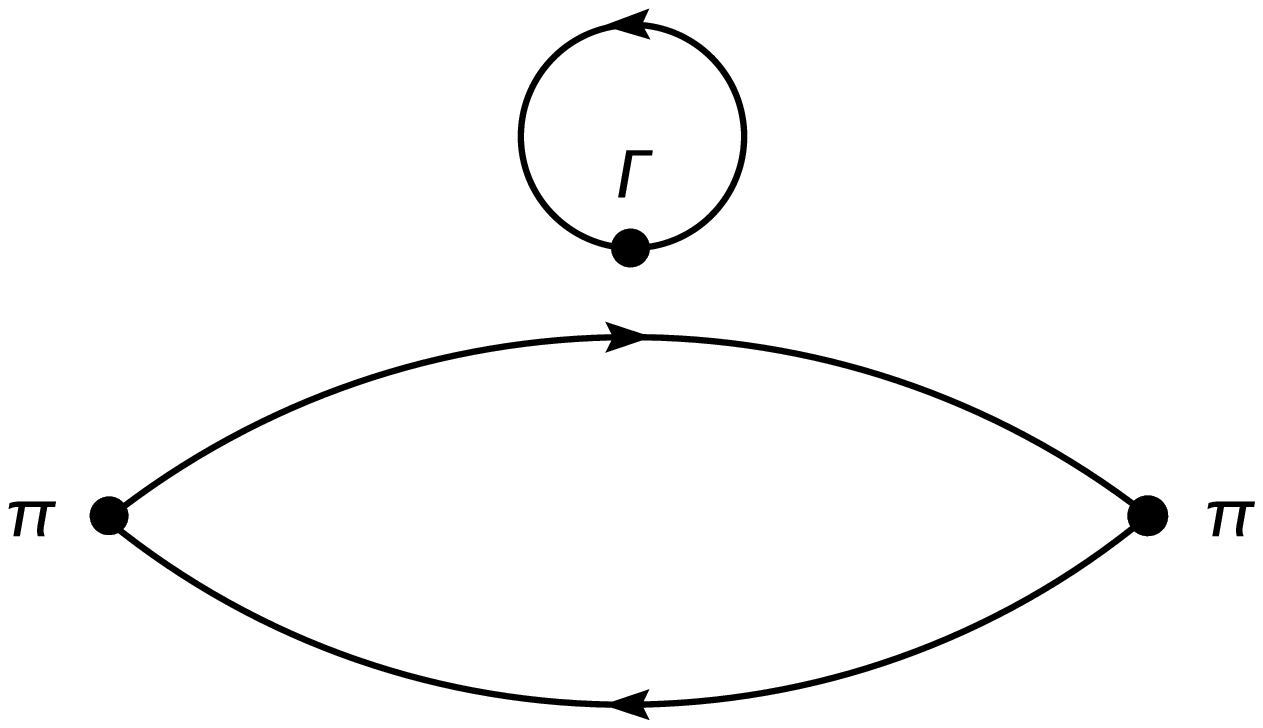}
\hspace{5mm}
\includegraphics[angle=0,width=0.3\linewidth,clip]%
                {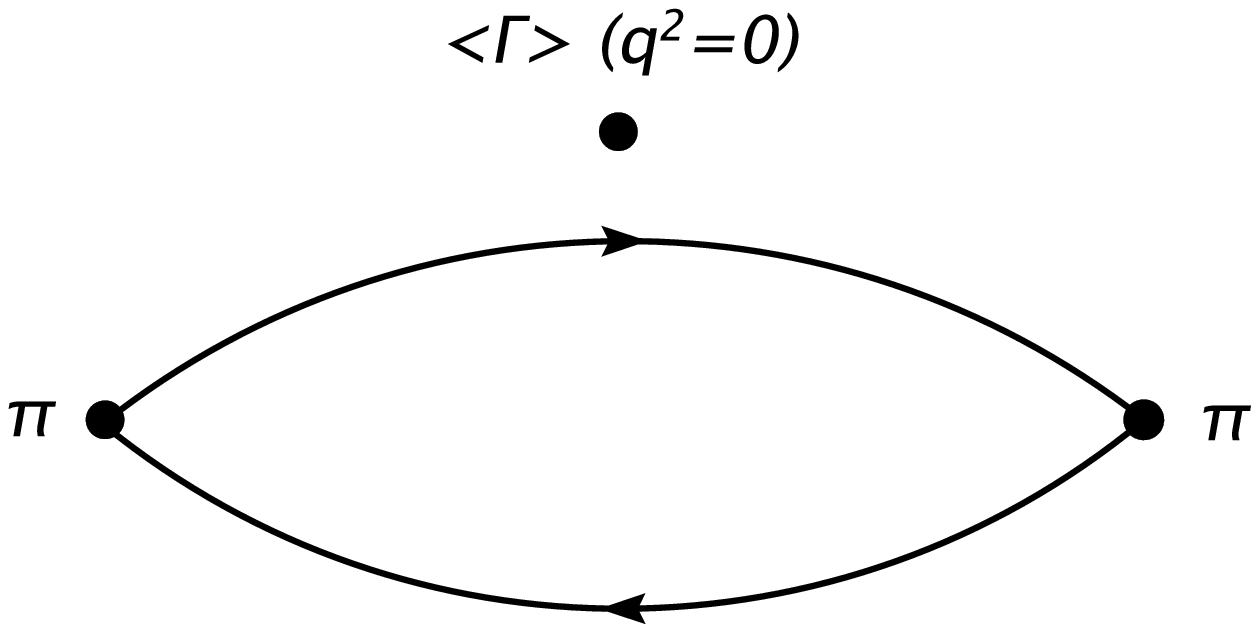}
\vspace{-2mm}
\caption{
   Connected (left-most diagram) and disconnected (middle diagram) 
   three point functions.
   Note that we have the contribution to $F_S(0)$
   from the right-most diagram 
   due to the non-zero vacuum expectation value of the scalar operator $S$.
}
\label{fig:meas:corr_3pt:diag}
\vspace{-4mm}
\end{center}
\end{figure}

From the $v$ and $w$ vectors,
we may construct meson fields at a temporal coordinate $t$ with
the Dirac matrix $\Gamma$ and spatial momentum $\bfp$ 
\bea
   \calO^{(k,l)}_{\Gamma,\phi}(t;\bfp)
   & = & 
   \sum_{\bfx,\bfr}
   \phi(\bfr)\, 
   w(\bfx+\bfr,t)^{(k)\dagger} \, 
   \Gamma \,
   v(\bfx,t)^{(l)}\,
   e^{-i \bfp \bfx}.
   \label{eqn:meas:meson_op}
\eea
For the smearing function $\phi(\bfr)$,
we choose the local $\phi_l(\bfr)\!=\!\delta_{{\bf r},{\bf 0}}$ and 
exponential function $\phi_{s}(\bfr)\!=\!\exp[-0.4|\bfr|]$.
Connected and disconnected three-point functions 
as well as the subtraction term of the vev contribution to the 
scalar form factor, shown in Fig.~\ref{fig:meas:corr_3pt:diag},
are calculated from these meson fields as 
\bea
    C_{\pi \Gamma \pi}^{\rm (conn)}
    (\Delta t, \Delta t^{\prime};\bfp,\bfp^\prime)
    & = & 
    \frac{1}{N_t} \sum_{t}
    \sum_{k,l,m=1}^{N_{\rm vec}}
    \calO^{(m,l)}_{\gamma_5,\phi_s}(t+\Delta t + \Delta t^\prime;\bfp^\prime) \, 
    \calO^{(l,k)}_{\Gamma,\phi_l}(t+ \Delta t;\bfp-\bfp^\prime)
    \nn \\[-2mm]
    & &
    \hspace{50mm}
    \times
    \calO^{(k,m)}_{\gamma_5,\phi_s}(t;-\bfp),
    \label{eqn:meas:msn_corr_3pt:conn}  
\eea
\bea
    C_{\pi \Gamma \pi}^{\rm (disc)}
    (\Delta t, \Delta t^\prime;\bfp,\bfp^\prime)
    & = & 
    \frac{1}{N_t} \sum_{t}
    \sum_{k,l=1}^{N_{\rm vec}}
    \calO^{(k,l)}_{\gamma_5,\phi_s}(t + \Delta t + \Delta t^\prime;\bfp^\prime) \, 
    \calO^{(l,k)}_{\gamma_5,\phi_s}(t;-\bfp)
    \nn \\[-2mm]
    & &
    \hspace{50mm}
    \times
    \sum_{m=1}^{N_{\rm vec}}
    \calO^{(m,m)}_{\Gamma,\phi_{l}}(t+\Delta t;\bfp-\bfp^\prime), 
    \label{eqn:meas:msn_corr_3pt:disc}  
    \\
    C_{\pi \Gamma \pi}^{\rm (vev)}
    (\Delta t, \Delta t^\prime;\bfp,\bfp^\prime)
    & = & 
    \frac{1}{N_t} \sum_{t}
    \sum_{k,l=1}^{N_{\rm vec}}
    \calO^{(k,l)}_{\gamma_5,\phi_s}(t+\Delta t + \Delta t^\prime;\bfp^\prime) \, 
    \calO^{(l,k)}_{\gamma_5,\phi_s}(t;-\bfp)
    \nn \\[-2mm]
    &&
    \hspace{35mm}
    \times 
    \left\langle
       \frac{1}{N_t} \sum_{t^\prime}
       \sum_{m=1}^{N_{\rm vec}}
       \calO^{(m,m)}_{\Gamma,\phi_{l}}(t^\prime;\bfp-\bfp^\prime)
    \right\rangle_{\rm conf},
    \label{eqn:meas:msn_corr_3pt:vev}  
\eea
where $\langle \cdots \rangle_{\rm conf}$ represents a Monte Carlo average.
We denote the temporal separation and spatial momentum for the initial 
(final) meson by $\Delta t$ and $\bfp$ ($\Delta t^\prime$ and $\bfp^\prime$),
respectively.

Our measurements are carried out at four values of 
the quark mass $m_{ud}$ 
in the range $290 \! \lesssim \! M_{\pi}~\mbox{[MeV]} \! \lesssim \! 520$.
We explore the region of the momentum transfer 
$-1.7 \! \lesssim \! q^2~\mbox{[GeV$^2$]} \! \leq \! 0$
by taking the meson momentum $\bfp$ with $|\bfp|\!\leq\!2$.
Note that the spatial meson momentum is shown in units of $2\pi a/L$ 
in this article.
While we have simulated only the trivial topological sector,
the effect of the fixed global topology is suppressed by the 
inverse of the space-time volume $\sim 1/V$ \cite{fixedQ}.

\section{Determination of pion form factors}

\begin{figure}[t]
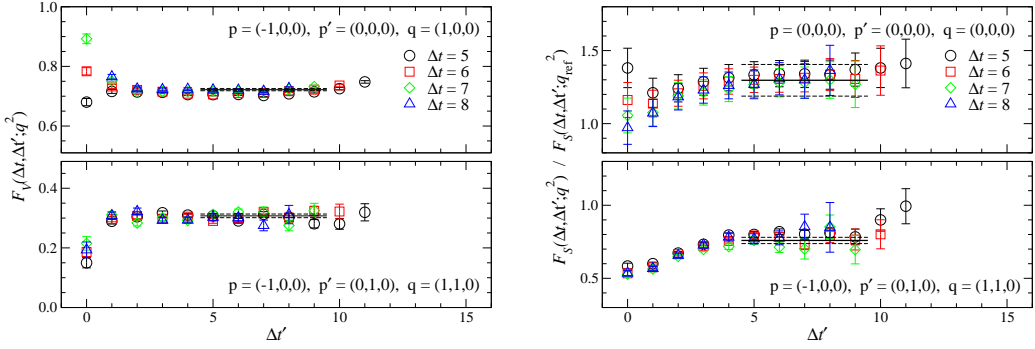

\begin{center}
\includegraphics[angle=0,width=0.43\linewidth,clip]%
                {pff_v_vs_dtsnk_m0050.eps}
\hspace{5mm}
\includegraphics[angle=0,width=0.43\linewidth,clip]%
                {pff_s_vs_dtsnk_m0050.eps}

\vspace{-3mm}
\caption{
   Effective value of $F_V(\Delta t, \Delta t^\prime;q^2)$ (left panels) and 
   $F_S(\Delta t, \Delta t^\prime; q^2)/F_S(\Delta t, \Delta t^\prime; q_{\rm ref}^2)$ 
   (right panels) at $m_{ud}\!\sim\!m_{s,\rm phys}/2$,
   where $m_{s,\rm phys}$ is the physical strange quark mass.
}
\label{fig:pff:pff_v_eff}
\end{center}
\vspace{-5mm}
\end{figure}

We calculate effective value of the vector form factor from a ratio 
\bea
   F_V(\Delta t,\Delta t^\prime;q^2)
   & = & 
   \frac{2\,M_\pi}{E_\pi(|\bfp|)+E_\pi(|\bfp^\prime|)}
   \frac{R_V(\Delta t,\Delta t^\prime; |\bfp|,|\bfp^\prime|,q^2)}
        {R_V(\Delta t,\Delta t^\prime; 0,0,0)},
   \label{eqn:pff:dratio:pff_v}
   \\
   R_V(\Delta t,\Delta t^\prime; |\bfp|,|\bfp^\prime|,q^2)
   & = &
   \frac{C_{\pi \gamma_4 \pi}^{\rm (conn)}
         (\Delta t,\Delta t^\prime; \bfp,\bfp^\prime)}
        {C_{\pi \pi, \phi_s \phi_l}(\Delta t;\bfp)\,
         C_{\pi \pi, \phi_l \phi_s}(\Delta t^\prime;\bfp^\prime)}.
   \label{eqn:pff:ratio:pff_v}
\eea       
Here $C_{\pi \pi, \phi \phi^\prime}$ is the pion two-point function
with the smearing function $\phi$ ($\phi^\prime$) for the source
(sink) operator, and it can also be calculated from the meson field 
Eq.~(\ref{eqn:meas:meson_op}).
We take the average of $R_V$ over momentum configurations 
corresponding to the same value of $q^2$.
This average as well as that over the location of the source operator 
in Eqs.~(\ref{eqn:meas:msn_corr_3pt:conn})\,--\,(\ref{eqn:meas:msn_corr_3pt:vev}) 
leads to an accurate estimate of $F_V(\Delta t,\Delta t^\prime;q^2)$
as shown in Fig.~\ref{fig:pff:pff_v_eff}.
The vector form factor $F_\pi(q^2)$ is determined from a constant 
fit in a range of $(\Delta t,\Delta t^\prime)$, 
where $F_V(\Delta t,\Delta t^\prime;q^2)$ shows a reasonable plateau.
We include the leading finite volume correction (FVC) \cite{FVC:PFF:V}
to $F_V(q^2)$.


The scalar form factor normalized at a certain momentum transfer 
$q_{\rm ref}^2$ can be calculated from a similar ratio 
\vspace{-1mm}
\bea
   \frac{F_S(\Delta t,\Delta t^\prime;q^2)}
        {F_S(\Delta t,\Delta t^\prime;q_{\rm ref}^2)}
   = 
   \frac{R_S(\Delta t,\Delta t^\prime; q^2)}
        {R_S(\Delta t,\Delta t^\prime; q_{\rm ref}^2)},
   \hspace{5mm}
   R_S(\Delta t,\Delta t^\prime; q^2)
   =
   \frac{C_{\pi 1 \pi}(\Delta t,\Delta t^\prime; \bfp,\bfp^\prime)}
        {C_{\pi \pi, \phi_s \phi_l}(\Delta t;\bfp)\,
         C_{\pi \pi, \phi_l \phi_s}(\Delta t^\prime;\bfp^\prime)},
   \label{eqn:pff:ratio:pff_s}
\eea
where
$C_{\pi 1 \pi} \! = \! C_{\pi 1 \pi}^{\rm (conn)} 
                  - C_{\pi 1 \pi}^{\rm (disc)} + C_{\pi 1 \pi}^{\rm (vev)}$.
As Fig.~\ref{fig:pff:pff_v_eff} indicates,
$F_S(q^2)$ at $q^2\!=\!0$ suffers from a relatively large 
statistical error than those at $q^2\!\ne\!0$ 
due to the severe cancellation between 
$C_{\pi 1 \pi}^{\rm (disc)}$ and $C_{\pi 1 \pi}^{\rm (vev)}$.
We therefore use $F_S(q^2)$ normalized at the smallest non-zero momentum 
transfer with $|{\bf q}_{\rm ref}|\!=\!1$ in the following analysis.
The normalized form factor $F_S(q^2)/F_S(q_{\rm ref}^2)$ is determined 
by a constant fit,
while FVC to $F_S(q^2)$ is not available so far 
and is not taken into account.

\section{Parametrization of $q^2$ dependence}

\begin{figure}[b]
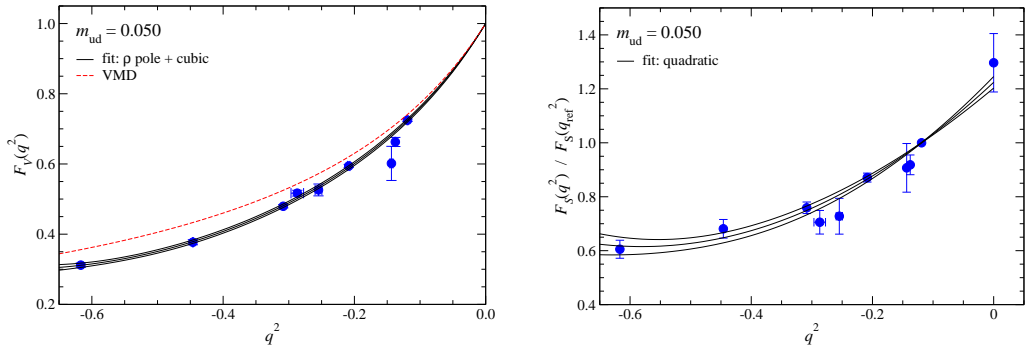

\begin{center}
\includegraphics[angle=0,width=0.43\linewidth,clip]{pff_v_vs_q2_m0050.eps}
\hspace{5mm}
\includegraphics[angle=0,width=0.42\linewidth,clip]{pff_s_vs_q2_m0050.eps}

\vspace{-3mm}
\caption{
   Vector (left panel) and normalized scalar form factors (right panel)
   at $m_{ud}\!\sim\!m_{s,\rm phys}/2$ as a function of $q^2$.
   Solid lines show our parametrization and its statistical error.
   In the left panel, 
   we also plot $\rho$ pole contribution expected from 
   the vector meson dominance hypothesis by the dashed line.
}
\label{fig:q2:pff_v_s}
\vspace{-7mm}
\end{center}
\end{figure}


The vector and scalar form factors are plotted as a function of $q^2$
in Fig.~\ref{fig:q2:pff_v_s}.
We observe that $F_V(q^2)$ is close to the pole dependence 
$1/(1-q^2/M_\rho^2)$ with $M_\rho$ measured at simulated $m_{ud}$.
%
%
Its $q^2$ dependence is therefore parametrized by the following form of 
the $\rho$ pole with a polynomial correction
to determine the charge radius $\crad_V$ and the curvature $c_V$
\bea
   F_V(q^2)
   & = & 
   \frac{1}{1-q^2/M_{\rho}^2} + c_1\,q^2 + c_2\,(q^2)^2 + c_3\,(q^2)^3
   \hspace{1mm} = \hspace{1mm}
   1 + \frac{1}{6} \crad_V \, q^2 + c_V \, (q^2)^2 + \cdots.
   \label{eqn:q2:vs_q2:pff_v}
\eea
Because the deviation from the $\rho$ pole is small,
we obtain a reasonable $\chi^2/{\rm dof}\!\sim\!1$,
and results for $\crad_V$ and $c_V$ are stable 
against the inclusion of the cubic correction term.

Such pole contribution in the scalar channel
is not clear within our statistical accuracy.
Our data can be fitted to a simple quadratic form
\bea
   F_S(q^2)
   & = & 
   F_S(0) \left( 1 + \frac{1}{6} \crad_S\, q^2 + c_S\, (q^2)^2 \right).
   \label{eqn:q2:vs_q2:pff_s}
\eea
We confirm that the result for the scalar radius $\crad_S$
is stable if we switch to the cubic or single pole form
$F_S(0)/(1-q^2/M_{\rm fit}^2)$ with $M_{\rm fit}$ as a fit parameter.
The curvature $c_S$ is however strongly depends on the choice of the 
parametrization form, and hence is not used in the following analysis.

\section{Chiral extrapolation}

\begin{figure}[t]
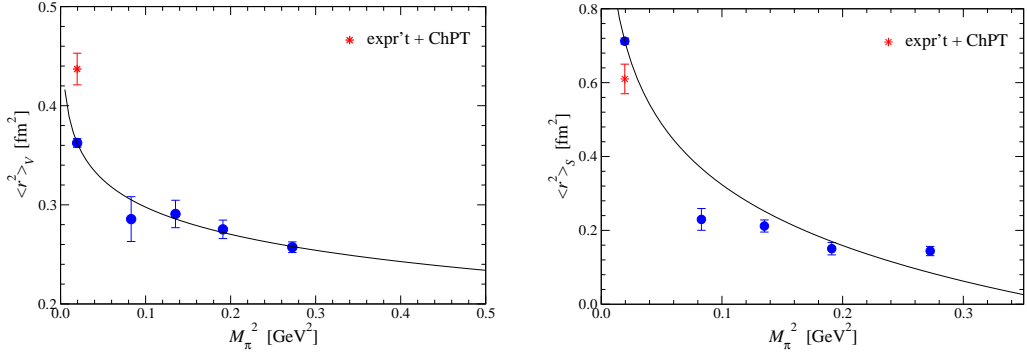

\begin{center}
\includegraphics[angle=0,width=0.43\linewidth,clip]{r2_v_vs_Mpi2.meas-pole+cubic.r2_v_nlo.eps}
\hspace{5mm}
\includegraphics[angle=0,width=0.42\linewidth,clip]{r2_s_vs_Mpi2.norm-mom0100.quad.r2_s_nlo.eps}

\vspace{-3mm}
\caption{
   Chiral extrapolation of $\crad_V$ (left panel) and $\crad_S$ (right panel)
   based on one-loop ChPT.
   Star symbols show the experimental value for $\crad_V$ \cite{PFF_V:ChPT:NNLO}
   and an indirect determination of $\crad_S$ through $\pi\pi$ scattering 
   \cite{PFF_S:ChPT:NNLO}.
}
\label{fig:chiral_fit:nlo}
\vspace{-5mm}
\end{center}
\end{figure}

In one-loop ChPT, 
the radii $\crad_V$ and $\crad_S$ are given by \cite{PFF:ChPT:NLO} 
\vspace{-1mm}
\bea
   \crad_V
   & = & 
   -\frac{1}{NF^2}\left( 1 + 6N\,l_6^r \right)
   -\frac{1}{NF^2}\ln\left[ \frac{M_\pi^2}{\mu^2} \right],
   \label{eqn:chiral_fit:r2_v:nlo}
   \\
   \crad_S
   & = &
    \frac{1}{NF^2}\left( -\frac{13}{2} + 6 N\,l_4^r \right)
   -\frac{6}{NF^2}\ln\left[ \frac{M_\pi^2}{\mu^2} \right],
   \label{eqn:chiral_fit:r2_s:nlo}
\eea
where $N\!=\!(4\pi)^2$.
We set the renormalization scale $\mu$ to $4\pi F$,
and fix $F$ to the value determined from our study of 
the pion mass and decay constant \cite{Spectrum:Nf2:RG+Ovr:JLQCD}.
%
These fits are however not quite successful 
as seen in Fig.~\ref{fig:chiral_fit:nlo}.
While our data of $\crad_V$ are fitted 
with reasonable $\chi^2/{\rm dof}\!\sim\!0.3$,
the value extrapolated to the physical quark mass $0.362(4)~\mbox{fm}^2$ 
is significantly smaller than experiment 
$0.437(16)~\mbox{fm}^2$ \cite{PFF_V:ChPT:NNLO}.
On the other hand, 
the one-loop formula for $\crad_S$ with the enhanced chiral log
fails to reproduce our data and results in large $\chi^2/{\rm dof}\!\sim\!16$.
We note that similar mild quark mass dependence of the radii is also observed 
by the ETM Collaboration with a different discretization on
a slightly finer lattice \cite{PFF:Nf2:Sym+tmW:ETMC}.
It is unlikely that 
the failure of the fits within one-loop ChPT
is caused by systematic uncertainties 
due to the fixed topology and the finite lattice spacing.

We then extend our analysis to two loops. 
The higher order contributions to the radii are given by~\cite{PFF_V:ChPT:NNLO,PFF_V:ChPT:NNLO:2}
\vspace{-1mm}
\bea
   \Delta \crad_V
   & = & 
    \frac{1}{N^2F^4}
    \left( \frac{13N}{192} - \frac{181}{48} + 6 N^2 r_{V,1}^r \right)\, M_\pi^2
   +\frac{1}{N^2F^4}
    \left( \frac{19}{6} - 12 N l_{1,2}^r \right)\, 
    M_\pi^2 \ln\left[ \frac{M_\pi^2}{\mu^2} \right],
   \label{eqn:chiral_fit:r2_v:nnlo}
\eea
\bea
   \Delta \crad_S
   & = & 
    \frac{1}{N^2F^4}
    \left( - \frac{23N}{192} + \frac{869}{108} 
           + 88 N l_{1,2}^r + 80 N l_2^r + 5 N l_3^r
           - 24 N^2 l_3^r l_4^r
           + 6 N^2 r_{S,1}^r \right)\, M_\pi^2
   \nn \\
   & & 
   +\frac{1}{N^2F^4}
    \left( - \frac{323}{36} + 124 N l_{1,2}^r
                            + 130 N l_2^r \right)\, 
    M_\pi^2 \ln\left[ \frac{M_\pi^2}{\mu^2} \right] 
   -\frac{65}{3N^2F^4}
    M_\pi^2 \ln\left[ \frac{M_\pi^2}{\mu^2} \right]^2.
   \label{eqn:chiral_fit:r2_s:nnlo}
\eea   
At two loops, the curvature $c_V$ has non-trivial contributions and 
can be included in our analysis
\bea
   c_V
   & = & 
    \frac{1}{60NF^2} \frac{1}{M_\pi^2}
   +\frac{1}{N^2F^4}
    \left( \frac{N}{720} - \frac{8429}{25920} 
                         + \frac{N}{3}l_{1,2}^r
                         + \frac{N}{6}l_6^r
                         + N^2 r_{V,2}^r \right) 
   \nn \\
   &   & 
   +\frac{1}{N^2F^4}
    \left( \frac{1}{108} + \frac{N}{3} l_{1,2}^r
                         + \frac{N}{6} l_6^r \right)\, 
    \ln\left[ \frac{M_\pi^2}{\mu^2} \right] 
   +\frac{1}{72N^2F^4}
    \ln\left[ \frac{M_\pi^2}{\mu^2} \right]^2.
   \label{eqn:chiral_fit:c_v:nnlo}
\eea
The analytic terms containing $r_{X,i}^r$ ($X\!=\!V,S$, $i\!=\!1,2$)
represent contributions from $O(p^6)$ chiral Lagrangian.
We denote the linear combination $l_1^r\!-\!l_2^r/2$
appearing commonly in $\crad_V$ and $c_V$ as $l_{1,2}^r$.

\begin{figure}[t]
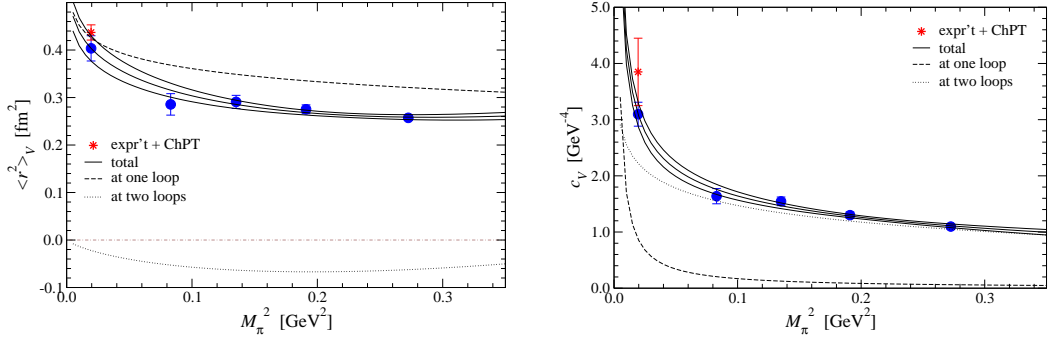

\begin{center}
\includegraphics[angle=0,width=0.44\linewidth,clip]{r2_v_vs_Mpi2.meas-pole+cubic.r2+c_v_nnlo.eps}
\hspace{5mm}
\includegraphics[angle=0,width=0.43\linewidth,clip]{c_v_vs_Mpi2.meas-pole+cubic.r2+c_v_nnlo.eps}
\vspace{-3mm}
\caption{
   Simultaneous chiral extrapolation of $\crad_V$ (left panel) and $c_V$ 
   (right panel) based on two-loop ChPT (solid lines). 
   We also plot one- and two-loop contributions 
   by dashed and dotted lines, respectively.
}
\label{fig:chiral_fit:r2_c_v:nnlo}
\vspace{-5mm}
\end{center}
\end{figure}

\FIGURE{
   \centering
   \includegraphics[angle=0,width=0.43\linewidth,clip]{r2_s_vs_Mpi2.meas-pole+cubic.r2_v_s_c_v_nnlo.eps}
   \vspace{-3mm}
   \caption{
      Chiral extrapolation of $\crad_S$ from simultaneous to
      $\crad_{V,S}$ and $c_V$ based on two-loop formulae.
   }
   \label{fig:chiral_fit:r2_v_s_c_v:nnlo}
}

While the two-loop formulae involve many LECs, 
the simultaneous fit to $\crad_V$ and $c_V$ has only four free parameters 
$l_6^r$, $l_{1,2}^r$, $r_{V,1}^r$ and $r_{V,2}^r$.
This fit plotted in Fig.~\ref{fig:chiral_fit:r2_c_v:nnlo}
shows that the two-loop contributions are significant 
in our simulated region of $m_{ud}$.
We obtain a reasonable value of $\chi^2/{\rm dof}\!\sim\!0.5$,
and the extrapolated values of $\crad_V$ and $c_V$ are
consistent with experiment \cite{PFF_V:ChPT:NNLO}.

The inclusion of $\crad_S$ into the simultaneous chiral fit introduces 
additional four free parameters.
In order to stabilize this fit,
we fix $l_2^r$ and $l_3^r$,
which appear only in the two-loop terms,
to a phenomenological estimate $\bar{l}_2\!=\!4.31$ 
\cite{PFF_S:ChPT:NNLO} and 
a lattice result $\bar{l}_3\!=\!3.44$ from our analysis 
of the pion spectroscopy \cite{Spectrum:Nf2:RG+Ovr:JLQCD}
\footnote{
The $\mu$ independent convention $\bar{l}_i$ is defined from 
$l_i^r\!=\!\gamma_i(\bar{l}_i+\ln[M_\pi^2/\mu^2])/2N$
with $\gamma_3\!=\!-1/2$, $\gamma_4\!=\!2$ and $\gamma_6\!=\!-1/3$.
}.
The extrapolation of $\crad_V$ and $c_V$ turns out to be 
consistent with those in Fig.~\ref{fig:chiral_fit:r2_c_v:nnlo}.
The extrapolation of $\crad_S$ is shown 
in Fig.\ref{fig:chiral_fit:r2_v_s_c_v:nnlo}.
From this simultaneous fit, we obtain 
\bea
   \crad_V 
   = 
   0.404(22)(22)~\mbox{fm}^2,
   \hspace{2mm}
   \crad_S
   = 
   0.578(69)(46)~\mbox{fm}^2,
   \hspace{2mm}
   c_V
   = 
   3.11(14)(86)~\mbox{GeV}^{-4}.
   \label{eqn:chiral_fit:r2_v_s_c_v:phys}
\eea  
The first error is statistical.
The second is systematic error 
estimated by changing the inputs for $l_2^r$ and $l_3^r$ 
to different phenomenological estimates in Ref.\cite{PFF_V:ChPT:NNLO},
and by limiting the fitting data to the radii ($\crad_V$ and $\crad_S$)
or those in the vector channel ($\crad_V$ and $c_V$).
We also test $\crad_S$ 
from the cubic parametrization for the $q^2$ dependence of $F_S(q^2)$.
Note that all the extrapolated values 
in Eq.~(\ref{eqn:chiral_fit:r2_v_s_c_v:phys}) are consistent with experiment.

We obtain
$\bar{l}_6 \! = \! 11.8(0.7)(1.3)$, 
$\bar{l}_4 \! = \! 4.06(44)(99)$,
and 
$l_{1,2}^r \! = \! -2.9(0.8)(2.4) \! \times \! 10^{-3}$
for the $O(p^4)$ LECs.
Our estimate of $\bar{l}_6$ is slightly smaller than 16.0(0.9)
obtained in Ref.\cite{PFF_V:ChPT:NNLO}
partly due to a deviation of $F$ 
between our lattice determination \cite{Spectrum:Nf2:RG+Ovr:JLQCD} 
and two-loop ChPT \cite{F:NhPT:NNLO}.
We note that $\bar{l}_4$ is consistent with our determination
$\bar{l}_4\!=\!4.12(56)$ from $F_\pi$  \cite{Spectrum:Nf2:RG+Ovr:JLQCD}
and a phenomenological estimate 4.39(22) \cite{PFF_S:ChPT:NNLO}.
Our results for the $O(p^6)$ LECs are 
$r_{V,1}^r \! = \! -1.1 \! \times \! 10^{-5}$,
$r_{V,2}^r \! = \! -4.0 \! \times \! 10^{-5}$
and $r_{S,1}^r \! = \! 1.3 \! \times \! 10^{-4}$
with substantial uncertainty of 50\,--\,100\,\%.


\section{Conclusions} 

In this article,
we report on our calculation of the pion form factors 
with two flavors of dynamical overlap quarks.
By employing the all-to-all quark propagators,
$F_S(q^2)$ is calculated including contributions from the disconnected 
diagrams for the first time.
The one-loop ChPT formulae fail to reproduce our data of $F_S(q^2)$.
In our analysis extended to two loops,
we observe significant two-loop contributions at our simulated quark masses,
and obtain $\crad_{V,S}$ and $c_V$ consistent with experiment.
Further investigations of systematics due to the fixed global 
topology and quenching of strange quarks are in progress 
by direct simulations in the non-trivial topological sectors 
and in three-flavor QCD.


\vspace{2mm}

Numerical simulations are performed on Hitachi SR11000 and 
IBM System Blue Gene Solution 
at High Energy Accelerator Research Organization (KEK) 
under a support of its Large Scale Simulation Program (No.~08-05).
This work is supported in part by the Grant-in-Aid of the
Ministry of Education (No.~18340075, 18740167, 19540286, 19740160,
20025010, 20039005, 20340047, and 20740156),
the National Science Council of Taiwan
(No.~NSC96-2112-M-002-020-MY3, NSC96-2112-M-001-017-MY3, NSC97-2119-M-002-001), 
and NTU-CQSE (No.~97R0066-69).


\end{document}